\documentclass[12pt]{article}
\usepackage{amsfonts,amssymb,amsmath,amsthm,amscd}
\usepackage{color}
\usepackage{hyperref}
\usepackage{verbatim} 

\newcommand {\calH}{\mbox{${\mathcal H}$}}
\newcommand {\calA}{\mbox{${\mathcal A}$}}
\newcommand {\calB}{\mbox{${\mathcal B}$}}

\newcommand {\calM}{\mbox{${\mathcal M}$}}
\newcommand {\calU}{\mbox{${\mathcal U}$}}

\newcommand{\sdirpr}{\rtimes}
\newtheorem{Definition}{Definition}
\newtheorem{Proposition}{Proposition}

\newtheorem{Theorem}{Theorem}

\renewcommand{\thesection}{\Roman{section}}

\newcommand{\x}{\times}
\newcommand{\C}{\mathbb{C}}              

\newcounter{mnotecount}[section]
\renewcommand{\themnotecount}{\thesection.\arabic{mnotecount}}
\newcommand{\mnote}[1]
{\protect{\stepcounter{mnotecount}}$^{\mbox{\footnotesize
$
\bullet$\themnotecount}}$ \marginpar{
\raggedright\tiny\em
$\!\!\!\!\!\!\,\bullet$\themnotecount: #1} }

\begin{document}

\title{Semidirect Product of Groupoids, Its Representations and Random Operators}
\author{Leszek Pysiak, \\
\url{lpysiak@mini.pw.edu.pl}\\
Technical University of Warsaw, \\
Plac Politechniki 1, 00-661 Warszawa, Poland \\
and Copernicus Center for Interdisciplinary Studies \\
ul. S{\l}awkowska 17, 31-016 Krak\'ow, Poland \\
\and Micha\l\ Eckstein \\
\url{michal.eckstein@uj.edu.pl}\\
Institute of Mathematics, Jagiellonian University, \\
ul. \L ojasiewicza 6, 30-348 Krak\'ow, Poland; \\
\and Michael Heller, \\
\url{mheller@wsd.tarnow.pl} \\
Copernicus Center for Interdisciplinary Studies \\
ul. S{\l}awkowska 17, 31-016 Krak\'ow, Poland \\
and Vatican Observatory, V-00120 Vatican City State;\\ 
\and
and Wies{\l}aw Sasin, \\
Technical University of
Warsaw,\\ Plac Politechniki 1, 00-661 Warszawa, Poland \\
and Copernicus Center for Interdisciplinary Studies \\
ul. S{\l}awkowska 17, 31-016 Krak\'ow, Poland \\
}

\date{\today}
\maketitle

\newpage

\begin{abstract}
One of pressing problems in mathematical physics is to find a generalized Poincar\'e symmetry that could be applied to nonflat space-times. As a step in this direction we define the semidirect product of groupoids $\Gamma_0 \rtimes \Gamma_1$ and investigate its properties. We also define the crossed product of a bundle of algebras with the groupoid $\Gamma_1$ and prove that it is isomorphic to the convolutive algebra of the groupoid $\Gamma_0 \rtimes \Gamma_1$. We show that families of unitary representations of semidirect product groupoids in a bundle of Hilbert spaces are random operators. An important example is the Poincar\'e groupoid defined as the semidirect product of the subgroupoid of generalized Lorentz transformations and the subgroupoid of generalized translations. 
\end{abstract}

\section{Introduction}
One of the main stumbling blocks in combining general relativity with quantum mechanics into a suitable generally relativistic quantum field theory is the fact that Poincar\'e group does not act on a curved space-time as a group of motions (e.g., \cite[p. 360]{Wald}). There are strong reasons to believe that to make a synthesis of general relativity and quantum physics would require a suitable generalization of the Poincar\'e group so as it would be able to enter into a fruitful interaction with geometry of curved manifolds (see, for instance, \cite{Chaichian,Fronsdal,Hollands}).
 
Natural generalization of group symmetries are groupoid symmetries. In the present work, we propose such a generalization. First, we define the semidirect product of groupoids $\Gamma_0 \rtimes \Gamma_1 $ where $\Gamma_0 $ and $\Gamma_1 $ are subgroupoids of the same groupoid $\Gamma $ (Sect. \ref{semidir}). Then, in Sect. \ref{algebras}, we define the action of the subgroupoid $\Gamma_1$ on a bundle $A$ of algebras related to the subgroupoid $\Gamma_0$. This allows us to define the crossed product of a bundle $A$ of algebras with $\Gamma_1$, $A \rtimes \Gamma_1$, which we prove to be isomorphic to the convolutive algebra of the groupoid $\Gamma_0 \rtimes \Gamma_1$. In \ref{reps} we show that families of unitary representations of the semidirect product groupoids in a bundle of Hilbert spaces form an algebra of random operators (in the sense of Connes \cite{Connes}), which can be completed to a von Neumann algebra. Finally, in Sect. \ref{Poincare}, we construct, as an example of the developed theory, the Poincar\'e groupoid being a semidirect product of the subgroupoid of generalized Lorentz transformations and the subgroupoid of generalized translations.

To fix notation we briefly describe the groupoid structure (for the definition see \cite[p.85]{SilvaWeinstein} or \cite[p. 269]{Landsman}).
A groupoid $\Gamma $ over $X$, or a groupoid with base  $X $, is a 7-tuple $(\Gamma, X , d, r, \epsilon , \iota , m) $ consisting of the following elements: (1)  sets $\Gamma $ and $X$, (2) surjections $(d, r) : \Gamma \rightarrow  X$, called source and target map,  respectively, (3) injection $\epsilon : X \rightarrow \Gamma $, $x \mapsto \epsilon(x), $ called  identity section or simply identity, (4)  map $ \iota : \Gamma  \rightarrow \Gamma $ , $\gamma \mapsto \iota (\gamma)=\gamma^{-1} $, called  inversion map. Moreover,  a composition law is defined, $ m : \Gamma^{(2)}
 \rightarrow \Gamma $, $( \gamma, \xi ) \mapsto m(\gamma,\xi)=\gamma \circ \xi $,
 with the domain $\Gamma^{(2)}:=\{(\gamma,\xi) \in \Gamma\times \Gamma : r(\xi )=d(\gamma)\}$ such that the following axioms are satisfied: (i) (associativity law) for arbitrary  $\gamma$,$\xi$,$\eta$ $\in \Gamma$ the triple product
 $(\gamma \circ \xi) \circ \eta $ is defined iff $\gamma \circ (\xi \circ \eta) $ is defined.  In such a case, we have $(\gamma \circ \xi  ) \circ \eta =\gamma \circ (\xi \circ \eta)$, (ii) (identities) for each $\gamma \in \Gamma$, $\epsilon(r(\gamma)) \circ \gamma = \gamma \circ \epsilon(d(\gamma)) = \gamma $, (iii) (inverses) for each $\gamma\in\Gamma , \gamma \circ \iota (\gamma) = \epsilon(r(\gamma))$, $\iota (\gamma)  \circ\gamma = \epsilon(d(\gamma))$. 
\par
For each $\gamma\in\Gamma$ the sets: $\Gamma^{x}=\{\gamma \in \Gamma : r(\gamma)=x ,x\in X \}$ and $ \Gamma_x=\{\gamma \in \Gamma : d(\gamma)=x  ,x \in X\}$ are defined. If $H \subset \Gamma $ and $H$ is closed under multiplication and inverses, $H$ can be naturally given the structure of the subgroupoid of $\Gamma $. A groupoid $\Gamma $ is called a Lie groupoid if the sets $\Gamma $ and $X$ are manifolds, the maps $d$ and $r$ are surjective submersions, and the structure maps are smooth \cite{SilvaWeinstein}. A groupoid $\Gamma $ is called transitive if for every pair $x, y \in X$ there exists $\gamma \in \Gamma$ such that $d(\gamma) = x$ and $r(\gamma) = y$.

\section{Semidirect Products of Groupoids}\label{semidir}
Given two groups one can obtain a new group in various ways. The resulting product group could be a direct product, a semidirect product, etc. A direct product of groupoids could be easily defined (see for instance \cite{SilvaWeinstein}). In this section we extend the concept of semidirect product to groupoids. Let $\Gamma $ be a topological, locally compact, and locally trivial groupoid over $X$ \cite{Paterson}. With these assumptions there exists the Haar system $\{\lambda_x\}_{x \in X}$ on $\Gamma $ \cite{Pysiak11}.

Let us recall the notion of an {\it isotropy groupoid} of the groupoid $\Gamma $
\[
\Gamma_0 := \bigcup_{x\in X}\Gamma^x_x
\]
where $\Gamma^x_x = \Gamma_x \cap \Gamma^x$. The subgroupoid $\Gamma_0$ is a wide subgroupoid of $\Gamma $, i.e., it is the subgroupoid of $\Gamma $ with the same base space

Let us now define another wide subgroupoid $\Gamma_1$ of $\Gamma $ such that $r \times d: \Gamma \rightarrow X \times X$, given by
\[
(r \times d)(\gamma ) = (r( \gamma ), d(\gamma )),
\]
is a surjection, i.e., the subgroupoid $\Gamma_1 $ is transitive. Here $d$ and $r$ are source and target mappings in $\Gamma $, respectively.

With the same symbol $\{\lambda_x\}_{x \in X}$ we denote the Haar system obtained by the restriction of measures $\lambda_x$ to the sets $\Gamma_{0,x}$. In the following, we write $d\lambda_x(\gamma_0) = d\gamma_0$ for the measure $\lambda_x$ on $\Gamma_{0,x}$. We also assume, for simplicity, that the groups $\Gamma_{0,x}$ are unimodular. It is straightforward to see that the restricted measures are right-invariant Haar measures on locally compact groups $\Gamma_{0,x}$ \cite{Pysiak11}.

\begin{Definition}
The {\em semidirect product} $\Gamma_0 \sdirpr \Gamma_1$ of the  groupoids $\Gamma_1$ and $\Gamma_0$ is the groupoid given by the following elements:
\begin{itemize}
\item
set
\[
\bar{\Gamma } = \Gamma_0 * \Gamma_1 = \{(\gamma_0, \gamma_1) \in \Gamma_0 \times \Gamma_1: d(\gamma_0) = r(\gamma_1)\},
\]
\item
source and target
\[\bar{d}(\gamma_0, \gamma_1) = d(\gamma_1)
\] \[
\bar{r}(\gamma_0, \gamma_1) = r(\gamma_0),
\]
\item
multiplication
\[
(\gamma_0, \gamma_1) \circ (\gamma '_0, \gamma '_1) = ( \gamma_0 \circ \alpha_{\gamma_1}(\gamma '_0), \gamma_1 \circ \gamma_1' ),
\]
provided that $\gamma_1$ and $\gamma'_1$ are composable; here we have introduced the following abbreviation (representing the action of $\Gamma_1$ on $\Gamma_0$)
\begin{align*}
\alpha_{\gamma_1}(\gamma_0) = \gamma_1 \circ \gamma_0 \circ \gamma_1^{-1},
\end{align*}
for $(\gamma_0,\gamma_1^{-1}) \in \bar{\Gamma }$; it can be easily seen that $\alpha_{\gamma_1}: \Gamma_{0,d(\gamma_1)} \rightarrow \Gamma_{0,r(\gamma_1)}$ is an isomorphism of groups.
\item
identity
\[
\epsilon_{\bar{\Gamma }}(x) = (\epsilon_{\Gamma_0}(x), \epsilon_{\Gamma_1}(x)) = (\epsilon(x), \epsilon(x))
\]
where $\epsilon_{\Gamma_0}$ and $\epsilon_{\Gamma_1}$ are identities of $\Gamma_0$ and $\Gamma_1$, respectively,
\item
inverse
\[
(\gamma_0, \gamma_1)^{-1} = (\alpha_{\gamma_1^{-1}}(\gamma_0^{-1}), \gamma_1^{-1}).
\]
\end{itemize}
\end{Definition}

Let $\gamma ,\, \gamma' \in \Gamma$. The elements $\gamma \sim \gamma'$ are equivalent (in the sense of equivalence relation) if there exists $\gamma_0 \in \Gamma_0$, with $d(\gamma_0 ) = r(\gamma )$, such that $\gamma_0 \circ \gamma = \gamma'$. On the set $\Gamma / \Gamma_0 = \{[\gamma ]_{\sim }: \gamma \in \Gamma \}$ we introduce the groupoid structure by defining
\[
\tilde{d}([\gamma ]) = d(\gamma ), \;\;\; \tilde{r}([\gamma ]) = r(\gamma ),
\] \[
[\gamma ] \circ [\gamma' ] = [\gamma \circ \gamma' ]
\]
provided that $d(\gamma) = r(\gamma ')$
\[ [\gamma ]^{-1} = [\gamma ^{-1}], \;\;\; \tilde{\epsilon} (x) = [\epsilon(x)].
\]

\begin{Proposition}\label{PropSemiDir}
The groupoid $\Gamma_1$ is isomorphic with the groupoid $\Gamma / \Gamma_0$ if and only if the semidirect product $\Gamma_0 \sdirpr \Gamma_1 $ is isomorphic with the groupoid $\Gamma $.
\end{Proposition}

\noindent
{\it Proof.} $\Rightarrow $) Let the isomorphism assumed in the Proposition be denoted by $j: \Gamma_1 \rightarrow \Gamma /\Gamma_0$. We define $J: \Gamma_0 \sdirpr \Gamma_1 \rightarrow \Gamma $ by
\[
J(\gamma_0, \gamma_1) = \gamma_0 \circ \gamma_1.
\]
First, we show that $J$ is a bijection. Let $\gamma \in \Gamma $ and $[\gamma ] \in \Gamma /\Gamma_0$. One has  $\gamma_1 = j^{-1}([\gamma]) \in \Gamma_1$ and let $\gamma_0 = \gamma \circ \gamma_1^{-1}$. From the isomorphism property of $j$ we have $d(\gamma_1) = \tilde{d}([\gamma]), \; r(\gamma_1) = \tilde{r}([\gamma ])$. But $\tilde{d}([\gamma]) = d(\gamma ), \; \tilde{r}([\gamma ]) = r(\gamma )$ and therefore
\[ d(\gamma_0) = d(\gamma_1^{-1}) = r(\gamma_1)
\] \[
r(\gamma_0 ) = r(\gamma ) = r(\gamma_1).
\]
Hence, $(\gamma_0, \gamma_1) \in \Gamma_0 \sdirpr \Gamma_1,\; (\gamma_0, \gamma_1) = J^{-1}(\gamma )$.

To show that $J$ is a homomorphism we apply $J$ to both sides of $(\gamma_0, \gamma_1) \circ (\gamma '_0, \gamma '_1) = (\gamma_0 \circ \alpha_{\gamma_1}(\gamma '_0), \gamma_1 \circ \gamma '_1)$ to obtain
\[ 
J((\gamma_0, \gamma_1) \circ (\gamma '_0, \gamma '_1)) = J(\gamma_0, \gamma_1) \circ J(\gamma '_0, \gamma '_1).\]

$\Leftarrow $) Let $I: \Gamma \rightarrow \Gamma_0 \sdirpr \Gamma_1$ be an isomorphism of groupoids. Then $I(\gamma ) = (\gamma_0, \gamma_1)$.
We define $i: \Gamma / \Gamma_0 \rightarrow \Gamma_1$ by
\[
i([\gamma ]) = \mathrm{pr}_2(I(\gamma )) = \gamma_1.
\]
Let us notice that if $I(\gamma ) = (\gamma_0, \gamma_1)$, i.e., $\gamma = \gamma_0 \circ \gamma_1$, then $\gamma_1 \in [\gamma ]$. Let now $\gamma' \in [\gamma ]$. This means that there exists $\bar{\gamma }_0 \in \Gamma_0$ such that
\[
\gamma' = \gamma'_0 \circ \gamma'_1 = \bar{\gamma}_0 \circ \gamma = \bar{\gamma}_0 \circ \gamma_0 \circ \gamma_1,
\]
which gives $(\gamma'_0, \gamma'_1) = (\bar{\gamma }_0 \circ \gamma_0, \gamma_1)$ under the action of the isomorphism $I$ and, in particular, $\gamma'_1 = \gamma_1$. This shows that the mapping $i$ is well defined.

The mapping $i$ is a surjection since, for $\gamma \in \Gamma_1$, one has $i([\gamma ]) = \gamma $, and it is also an injection since $(i^{-1})(\gamma_1) = [\gamma_1]$. Being a homomorphism it is, therefore, an isomorphism. \qed

\section{Algebras Associated with the Semidirect Product of Groupoids}\label{algebras}
In this section we define the action of the subgroupoid $\Gamma_1$ on a bundle $A$ of algebras related to the subgroupoid $\Gamma_0$. The action, denoted by $A \rtimes \Gamma_1$, is called the crossed product of a bundle $A$ of algebras with $\Gamma_1$. We also prove that $A \rtimes \Gamma_1$ is isomorphic to the convolutive algebra of the groupoid $\Gamma_0 \rtimes \Gamma_1$.

With a semidirect product of groupoids we can associate various algebraic structures. Let us first define the bundle $A$ of algebras $(A_x, \bullet )_{x \in X}$ where $A_x = L^1(\Gamma_{0,x})$ with convolution
\[
(a_1 \bullet a_2)(\gamma_0) = \int_{\Gamma_{0,x}}a_1(\gamma'_0)a_2(\gamma'^{-1}_{0} \circ \gamma_0)d\gamma'_0.
\]

The action $\alpha : \Gamma_1 \x \Gamma_0 \to \Gamma_0$, as it is defined in section \ref{semidir}, induces the dual action $\beta : \Gamma_1 \x A \to A$ given by
\begin{align*}
\beta_{\gamma_1} : A_{r(\gamma_1)} \to A_{d(\gamma_1)}, && (\beta_{\gamma_1}(a))(\gamma_0) = a( \alpha_{\gamma_1}(\gamma_0) ).
\end{align*}
This allows us to formulate the following definition

\begin{Definition}
The {\em crossed product} of the bundle of algebras $A = (A_x, \bullet )_{x \in X}$ and the groupoid $\Gamma_1$ is the algebra
\begin{align*}
 \calB = A \rtimes \Gamma_1 = \{ F \in L^1(\Gamma_1, A): F(\gamma_1) \in A_{r(\gamma_1)}, \forall \gamma_1 \in \Gamma_1 \},
\end{align*} 
with twisted convolution as multiplication
\[
(F_1 \circledast F_2)(\gamma_1) = \int_{\Gamma_1^{r(\gamma_1)}}F_1(\gamma_1') \bullet \beta_{\gamma_1'^{-1}}(F_2(\gamma_1'^{-1} \circ \gamma_1))d\gamma_1'.
\]
\end{Definition}
By $d\gamma_1' $ we denote the restriction of the suitable Haar measure $\lambda_x$ on $\Gamma$ to $\Gamma_1^{r(\gamma_1)}$.

We also define the algebra $\calA = (L^1(\Gamma_0 \rtimes \Gamma_1, \C), *)$. The fact that functions $f_1, f_2 \in \calA $ are integrable allows us to define their convolution in the following way
\[
(f_1 * f_2)(\gamma_0,\gamma_1) = \int_{\Gamma_{1}^{r(\gamma_1)}} \int_{\Gamma_{0,r(\gamma_1)}} f_1(\gamma'_0, \gamma'_1) f_2 \big( (\gamma'_0, \gamma'_1)^{-1} \circ (\gamma_0, \gamma_1) \big) d\gamma'_0 d\gamma'_1.
\]

\begin{Theorem}
\label{Iso}
The algebra $\calB $ is isomorphic with the algebra $\calA $; the isomorphism $K: \calB \rightarrow \calA$ is given by
\begin{align*}
(KF)(\gamma_0, \gamma_1) = (F(\gamma_1))(\gamma_0), && \text{for all } (\gamma_0, \gamma_1) \in \bar{\Gamma} 
\end{align*}
\end{Theorem}

\textit{Proof.} First, we show that $K$ is an homomorphism of algebras. Indeed,
\begin{multline*}
(KF_1 * KF_2)(\gamma_0, \gamma_1) =  \\
\shoveleft{ \quad = \int_{\Gamma_{1}^{r(\gamma_1)}} \int_{\Gamma_{0,r(\gamma_1)}} (KF_1)(\gamma'_0, \gamma'_1) (KF_2) \big( (\gamma'_0, \gamma'_1)^{-1} \circ (\gamma_0, \gamma_1) \big) d\gamma'_0 d\gamma'_1 = }\\
\shoveleft{ \quad = \int_{\Gamma_{1}^{r(\gamma_1)}} \int_{\Gamma_{0,r(\gamma_1)}} (KF_1)(\gamma'_0, \gamma'_1) (KF_2) \big( \alpha_{\gamma_1'^{-1}}(\gamma_0'^{-1} \circ \gamma_0), \gamma_1'^{-1} \circ \gamma_1) d\gamma'_0 d\gamma'_1 = }\\
\shoveleft{ \quad = \int_{\Gamma_{1}^{r(\gamma_1)}} \int_{\Gamma_{0,r(\gamma_1)}} (F_1(\gamma_1'))(\gamma_0') (F_2(\gamma_1'^{-1} \circ \gamma_1)) \big( \alpha_{\gamma_1'^{-1}}(\gamma_0'^{-1} \circ \gamma_0) \big) d\gamma'_0 d\gamma'_1 = }\\
\shoveleft{ \quad = \int_{\Gamma_{1}^{r(\gamma_1)}} \int_{\Gamma_{0,r(\gamma_1)}} (F_1(\gamma_1'))(\gamma_0') \big( \beta_{\gamma_1'^{-1}} F_2(\gamma_1'^{-1} \circ \gamma_1) \big) (\gamma_0'^{-1} \circ \gamma_0) d\gamma'_0 d\gamma'_1 = }\\
= \big( (F_1 \circledast F_2) (\gamma_1) \big) (\gamma_0).
\end{multline*}
We have taken into account that
\begin{flalign*}
(\gamma'_0, \gamma'_1)^{-1} \circ (\gamma_0, \gamma_1) & = (\gamma_1'^{-1} \circ \gamma_0'^{-1} \circ \gamma_1', \gamma_1'^{-1}) \circ (\gamma_0, \gamma_1) = \\
& = (\gamma_1'^{-1} \circ \gamma_0'^{-1} \circ \gamma_1' \circ \alpha_{\gamma_1'^{-1}}(\gamma_0), \gamma_1'^{-1} \circ \gamma_1) = \\
& = (\alpha_{\gamma_1'^{-1}}(\gamma_0'^{-1} \circ \gamma_0), \gamma_1'^{-1} \circ \gamma_1).
\end{flalign*}

It remains to show that $K$ is a bijection. Let $F \in L^1(\Gamma_1, A)$ and suppose that $KF = 0$, i.e.,
\[
(KF)(\gamma_0, \gamma_1) = (F(\gamma_1))(\gamma_0) = 0
\]
for almost all $(\gamma_0,\gamma_1) \in \bar{\Gamma}$ in the sense of $L^1$. Therefore, $F(\gamma_1) = 0$ for almost all $\gamma_1 \in \Gamma_1$, and $(F(\gamma_1))(\gamma_0) = 0$ for almost all $\gamma_0 \in \Gamma_{0, r(\gamma_1)}$. Thus $F = 0$ which shows that $K$ is an injection.

It is also a surjection since for every $f \in L^1(\Gamma_0 \rtimes \Gamma_1, \C)$ there exists $F \in L^1(\Gamma_1, A)$, namely $(F(\gamma_1))(\gamma_0) = f(\gamma_0, \gamma_1)$. \qed

\section{Unitary Representations and Random Operators}\label{reps}
Let us now investigate the unitary representations of groupoids and the associated algebras.
\begin{Definition} \cite{Paterson} Unitary representation of a groupoid $\Gamma $ over $X$ is a pair $(\calU, \calH )$ where \calH \ is a bundle of Hilbert spaces over $X$, $\calH = \{H_x\}_{x \in X}$, and $\calU = \{\calU (\gamma ): \gamma\in\Gamma\}$ is a family of unitary transformations $\calU(\gamma ): H_{d(\gamma )} \rightarrow H_{r(\gamma )}$ such that
\begin{enumerate}
\item
$\calU(\epsilon(x)) = \mathrm{id}_{H{_x}}$ for $x \in X$,
\item
$\calU(\gamma_1 \circ \gamma_2) = \calU(\gamma_1) \calU(\gamma_2)$ for $(\gamma_1, \gamma_2) \in \Gamma^{(2)}$,
\item
$\calU(\gamma^{-1}) = \calU(\gamma )^{-1}$ for almost all $\gamma \in \Gamma $ with respect to the measure as it is defined in \cite[p. 92]{Paterson},
\item
for every $\varphi, \psi \in L^2(X, \calH )$ the function
\[
\Gamma \ni \gamma \mapsto \big( \calU(\gamma ) \varphi(d(\gamma )), \psi (r(\gamma ) \big)_{r(\gamma )} \in \C
\]
is measurable on $\Gamma $.
\end{enumerate}
\end{Definition}

Let now $(\calU_0 , \calH)$ be a unitary representation of the isotropy groupoid $\Gamma_0$ in a Hilbert bundle $\calH = \{H_x\}_{x \in X}$ over $X$, and $\{i_x^y\}_{x,y \in X}$ a family of isomorphisms of Hilbert spaces, $i_x^y: H_x \rightarrow H_y$.

\begin{Definition}
The simple extension of the representation $(\calU_0 , \calH )$ of the groupoid $\Gamma_0 $ to the groupoid $\Gamma = \Gamma_0 \sdirpr \Gamma_1$ is the representation  $(\calU , \cal H )$ of $\Gamma $ given by
\[
\calU (\gamma_0, \gamma_1) = \calU_0(\gamma_0 ) \circ i_x^y
\]
for $\gamma_0 \in \Gamma_{0,y}, \gamma_1 \in \Gamma_{1,x}$ such that $x = d(\gamma_1 ), y = r(\gamma_1)$.
\end{Definition}
It is easy to see that the family $\{i_x^y\}_{x,y \in X}$ determines the unitary representation of the groupoid $\Gamma_1$ in the Hilbert bundle $\{H_x \}_{x \in X}$.

Let us denote $\calU_1(\gamma_1)h_x = i_x^y(h_x),\, h_x \in H_x$ for $\gamma_1 \in (\Gamma_1)_x^y$, and let us notice that the following commutation relation is satisfied
\[
\calU_1(\gamma_1) \calU_0(\gamma_0)\calU_1(\gamma_1^{-1}) = \calU_0(\alpha_{\gamma_1}(\gamma_0))
\]
for $\gamma_1, \gamma_0$ such that $d(\gamma_1) = d(\gamma_0)$.

We now recall the definition of a random operator \cite{Connes}.
\begin{Definition}
A \textit{random operator} in a Hilbert bundle $\mathcal{H} = \{H_x\}_{x \in M}$ is defined to be a map
\[
M \ni x \mapsto b_x \in {\cal B}(H_x),
\]
where $\calB(H_x)$ denotes the algebra of bounded operators on $H_x$. It must satisfy the following conditions:
\begin{enumerate}
\item
if $\{\psi_i\}_{i=1}^{\infty }$ is a measurable field of bases in $\calH $ then the function
\[
M \ni x \mapsto (B_x \psi_i(x), \psi_j(x))_{x}, \quad \text{for } \, i,j = 1,2, \ldots
\]
is measurable,
\item
the family of operators $B_x,\, x \in M$, is essentially bounded in the operator norm, i.e., $\mathrm{ess \ sup}_{x \in X} ||B_x|| < \infty $.
\end{enumerate}
\end{Definition}

For $a \in L^1(\Gamma_0)$ we define the operators $\calU_{0,x}(a): H_x \rightarrow H_x$ by
\[
\calU_{0,x}(a)(h_x) := \int_{\Gamma_{0,x}}a(\gamma )\calU_0(\gamma ) h_x d\lambda_x
\]
for $h_x \in H_x$; $\{\lambda_x\}_{x \in X}$ is here the Haar system on the groupoid $\Gamma_0$ defined by the Haar measure on the group $G$.

\begin{Proposition}
The family of operators $r_a := \{\calU_{0,x}(a)\}_{x \in X}$ forms a random operator in the Hilbert bundle \calH .
\end{Proposition}

\noindent
\textit{Proof.} Since $a \in L^1(\Gamma_0)$, it is enough to prove that the condition (2) is satisfied. Indeed,
\begin{align*}
||\calU_{0,x}(a)|| & = \sup_{||h_x||\leq 1} ||\calU_{0,x}(a)h_x|| = \sup_{||h_x||\leq 1} ||\int_{\Gamma_{0,x}} a(\gamma )\calU_0 (\gamma ) h_x d \lambda_x || \leq \\
& \leq \sup_{||h_x||\leq 1} \int_{\Gamma_{0,x}} |a(\gamma )|\, ||\calU_0 (\gamma ) h_x || d \lambda_x  = \sup_{||h_x||\leq 1} \int_{\Gamma_{0,x}} |a(\gamma )|\, || h_x || d \lambda_x  \leq \\
& \leq \int_{\Gamma_{0,x}} |a(\gamma )|d \lambda_x  \leq  \mathrm{vol} (\mathrm{supp} \, a) \cdot \sup |a(\gamma )| < \infty . \;
\end{align*} \qed

Let us notice that for $\gamma_0 \in \Gamma_{0,x}, \, \gamma \in \Gamma_x^y$ we have $\alpha_{\gamma}(\gamma_0 ) \in \Gamma_{0,y}$. Let us also define
\[
(\alpha^*_{\gamma}(a))(\gamma_0) = a(\alpha_{\gamma}(\gamma_0)))
\]
for $a \in L^1(\Gamma_0) $ and $\gamma_0 \in \Gamma_0$, $\gamma \in \Gamma_x^y$. 

The following transformation rules for the operators $r_a$ hold
\begin{align*}
\calU_0(\gamma_0) \calU_{0,x}(a)\calU_0(\gamma_0^{-1}) & = \calU_{0,x}(\alpha^*_{\gamma_0}(a)), \\
\calU_1(\gamma_1) \calU_{0,x}(a)\calU_1(\gamma_1^{-1}) & = \calU_{0,y}(\alpha^*_{\gamma_1}(a)).
\end{align*}

If an operator $r_a, \, a \in L^1(\Gamma_0) $, satisfies the above conditions, it is said to be equivariant. Let us notice, that the second rule allows us to call the representation of $L^1(\Gamma_0)$ in $\calH$ a \textit{covariant representation} with respect to the induced groupoid action $\alpha^*$.

Let $\calM_0$ denote the algebra of operators of the form $r_a, \, a \in L^1(\Gamma_0) $. Then the algebra $\calM = (\calM_0)''$ is a von Neumann algebra. The bicommutant $(\calM_0)''$ is considered here in the Hilbert space $\int_{\oplus }H_x d\mu (x)$. We call $\calM$ the von Neumann algebra of the groupoid $\Gamma_0$. In a forthcoming publication we will describe the von Neumann algebra of the semidirect product of groupoids $\Gamma_0 \rtimes \Gamma_1$.

\section{An Example -- the Poincar\'e Groupoid}\label{Poincare}
In this section we consider a special case of the above constructions which may have direct applications to physics. Let $E$ be a frame bundle over $M$ (we can think of $M$ as of space-time). We define
\[
\tilde{\Gamma} = \{(p_1,p_2): p_1, p_2 \in E\},
\]
and introduce the following equivalence relation in $\tilde{\Gamma }$
\[
(p_1,p_2) \sim (p'_1, p'_2) \Leftrightarrow \exists g \in G,\, p'_1 = p_1g, p'_2 = p_2g.
\]
To simplify notation, the equivalence class of the element $(p_1, p_2)$ will be denoted by $[p_1,p_2]$. We also denote $\tilde{\Gamma }/\!\!\sim \;
 = \Gamma = E \times_G E$. We introduce the groupoid structure in $\Gamma $ in the following way
\begin{itemize}
\item
composition  $[p_1,p_2] \circ [p_3,p_4] = [p_1, p_4g^{-1}]$ is defined only if there exists $g \in G$ such that $p_3= p_2g$,
\item
source and target
\[
d([p_1,p_2]) = \pi_M(p_2),
\]
\[ r([p_1,p_2]) = \pi_M(p_1),
\]
\item
identity $\epsilon (x) = [p,p],\, \pi_M(p) = x$,
\item
inverse $[p_1,p_2]^{-1} = [p_2,p_1]$.
\end{itemize}
The following sets are naturally defined
\[
\Gamma^x = \{[p_1,p_2]: \pi_M(p_1) = x\},
\]\[
\Gamma_y = \{[p_1,p_2]: \pi_M(p_2) = y\}.
\]
In the literature groupoid $\Gamma $ is called {\it gauge groupoid}. It is a transitive groupoid. Indeed, for any $x, y \in M$ one finds $[p_1, p_2] \in \Gamma $ such that $\pi_M(p_1) = x, \, \pi_M(p_2) = y$.

We construct two subgroupoids of $\Gamma $: the subgroupoid $\Gamma_0 = \{[p_1,p_1g]: p_1 \in E, g\in G\}$ of the gauge groupoid $\Gamma $ consists of equivalence classes of Lorentz transformations
\[
(p_1, p_1g) \sim (p_1g_0, p_1gg_0),
\]
and the subgroupoid $ \Gamma_1 = \{[s(x), s(y)]: x, y \in M\}$, where $s(x)$ is a cross section $s: M \rightarrow E$, consisting of generalized translations in $M$. In fact, there is an isomorphism between $\Gamma_1$ and  $M \times M$ for each $s$.

Let us notice that $\Gamma_0$ is the isotropy groupoid of the groupoid $\Gamma $, i.e., $\Gamma_0 = \{\Gamma^x_x\}_{x\in M}$ where $\Gamma^x_x = \Gamma_x \cap \Gamma^x$.

\begin{Proposition}
The semidirect product $\Gamma_0 \sdirpr \Gamma_1$, for any cross section $s$, is isomorphic with the gauge groupoid $\Gamma $.
\end{Proposition}

{\it Proof .} It is enough to show that there exists an isomorphism $\Gamma_1 \simeq \Gamma /\Gamma_0$. Let $\rho : \Gamma \rightarrow \Gamma /\Gamma_0$ be a canonical projection; it is a homomorphism of groupoids. Let further $i: \Gamma /\Gamma_0 \rightarrow \Gamma_1$ be as defined in the proof of Proposition 1. We can write
\[i \big( \rho ([s(x)g, s(y)]) \big) = i \big( \rho ([s(x)g, s(x)] \circ [s(x), s(y)]) \big) = [s(x), s(y)].
\]
Therefore, $i$ is a homomorphism of groupoids. It is also a bijection. Indeed, $i$ is obviously a surjection on $\Gamma_1$, and 
\[
i^{-1}([s(x), s(y)]) = \rho([s(x), s(y)]
\]
shows that it is an injection. \qed

Let us notice that translations in $E$ depend on the cross section $s$; this can be regarded as a kind of gauge fixing. If $G$ is the Lorentz group, the above proposition justifies calling the gauge groupoid the Poincar\'e groupoid (as a semidirect product of generalized Lorentz transformations and generalized translations).

For the Poincar\'e groupoid the Haar system $\{\lambda_x\}_{x \in M}$ on $\Gamma_{0,x}$ is of the form
\[
\int_{\Gamma_{0,x}}f(\gamma_0)d\lambda_x(\gamma_0) = \int_G f([p_0, p_0g])dg
\]
where $G$ is the Lorentz group, $dg$ is a Haar measure on $G$ and $p_0$ is a selected element of $E_x$. On $\Gamma_1$ one has
\[
\int_{\Gamma_{1,y}} f(\gamma_1) d\lambda_y = \int_M f([s(x), s(x)]) d\mu (x)
\]
where $\mu $ is a Lebesgue measure on $M$.

We have the bundle of algebras $A = \{A_x\}_{x \in M}$ where $A_x = L^1(\Gamma_{0,x})$ and, on the strength of Theorem \ref{Iso}, the crossed product $\calB = A \rtimes \Gamma_1$ is isomorphic with the algebra $(\calA , *)$ associated with the Poincar\'e groupoid $\Gamma = \Gamma_0 \rtimes \Gamma_1$. Let us write down the product in $\calA$ explicitly.

Let $f_1, f_2 \in \calA $ and $[s(x)g, s(y)] \in \Gamma = \Gamma_0 \rtimes \Gamma_1$, then we have
\begin{multline*}
(f_1 * f_2)([s(x)g, s(x)], [s(x), s(y)]) = \\
\shoveleft{ \quad = \int_{\Gamma_1^x} \int_{\Gamma_{0,x}}f_1([s(x)g',s(x)],[s(x),s(z)]) \times} \\
\times f_2(([s(x)g',s(x)],[s(x),s(z)])^{-1} \circ ([s(x)g,s(x)],[s(x),s(y)])dg' d\mu(z) = \\
\shoveleft{ \quad = \int_{\Gamma_1^x} \int_{\Gamma_{0,x}}f_1([s(x)g', s(x)],[s(x),s(z)]) \times} \\
\times f_2([s(z)g'^{-1}g,s(z)],[s(z),s(y)])dg'd\mu (z).
\end{multline*}
In this formula the nonlocal character of the generalized Poincar\'e symmetry (of the groupoid type) can explicitly be seen.

\end{document}